\documentclass{IEEEtran}
\usepackage{algorithm,algorithmic,amsmath,color,comment,float,graphicx,latexsym,subfigure,tikz,times,url}

\pdfoutput=1

\title{Assessment of Effectiveness of Content Models for Approximating Twitter Social Connection Structures}

\author{\IEEEauthorblockN{
Kuntal Dey\IEEEauthorrefmark{1}\IEEEauthorrefmark{2}, 
Sahil Agrawal\IEEEauthorrefmark{2},
Rahul Malviya\IEEEauthorrefmark{2} and
Saroj Kaushik\IEEEauthorrefmark{2}}\\
\IEEEauthorblockA{\IEEEauthorrefmark{1}IBM Research India, New Delhi, India}\\
\IEEEauthorblockA{\IEEEauthorrefmark{2}Indian Institute of Technology, New Delhi, India}\\
}

\begin{document}

\maketitle

\begin{abstract}
This paper explores the social quality (goodness) of community structures formed across Twitter users, where social links within the structures are estimated based upon semantic properties of user-generated content (corpus).
We examined the overlap of the community structures of the constructed graphs, and followership-based social communities, to find the social goodness of the links constructed.
Unigram, bigram and LDA content models were empirically investigated for evaluation of effectiveness, as approximators of underlying social graphs, such that they maintain the {\it community} social property.
Impact of content at varying granularities, for the purpose of predicting links while retaining the social community structures, was investigated.
$100$ discussion topics, spanning over $10$ Twitter events, were used for experiments.
The unigram language model performed the best, indicating strong similarity of word usage within deeply connected social communities.
This observation agrees with the phenomenon of evolution of word usage behavior, that transform individuals belonging to the same community tending to choose the same words, made by \cite{danescu2013no}, and raises a question on the literature that use, without validation, LDA for content-based social link prediction over other content models.
Also, semantically finer-grained content was observed to be more effective compared to coarser-grained content.
\end{abstract}

\section{Introduction}
\label{sec:intro}
User generated content on social networks such as Facebook, and microblogs such as Twitter, has become a trending research topic in recent years.
Microblogs have been studied from a number of research perspectives, such as information diffusion \cite{kwak2010twitter} and spread of ideas \cite{tsur2012s}.

\subsection{Motivation}
\label{sec:motivation}
Several works, such as \cite{adamic2003friends}, \cite{jeh2002simrank}, \cite{katz1953new}, \cite{liben2007link} and \cite{salton1986introduction} address predicting (constructing) social links between pairs of users, from graph attributes.
\cite{dong2012link} and \cite{yin2011structural} focus on graph structure and properties.
\cite{puniyani2010social} predicts links based upon semantic content.
\cite{ritter2010unsupervised} studies language-based conversation modeling of Twitter users.

Statistical count of correctness of predicted links does not reveal any structural or social insight about the prediction.
Two cases of predicting links using two independent algorithms could have similar precision and recall, but could form completely different graph (social) structures.
This would imply radically different community structures and dynamics.

The goodness of the predicted social links is measured in literature, including \cite{puniyani2010social}, using statistical phenomenon such as accuracy, precision and recall, not {\it social} attributes.
\cite{puniyani2010social} selects Latent Dirichlet Allocation (LDA) \cite{blei2003latent} without exploring other language models.
Also, no work investigates the impact of different user-generated content (corpus) granularities.
This necessitates our work.

\subsection{Contributions of Our Work}
\label{sec:contributions}
In the current work, we overlay the community structures formed by two graphs: a content graph and a social friendship graph.
Content graph is created by collecting the lifetime content generated (tweets made) by the user, related to a given event.
The links between user pairs, are constructed (predicted) based upon the content generation similarity between user pairs, as found by content models, such as unigram, bigram and LDA.
We use the popular modularity \cite{newman2006modularity} maximization technique, which is clearly by far the most widely accepted definition of social network communities \cite{nicosia2009extending}.
We specifically apply the BGLL \cite{blondel2008fast} algorithm implementing the technique, to identify the structural communities implicitly present in the constructed graphs (predicted links).
Community overlap of each pair of graphs is quantified using normalized mutual information (NMI) \cite{coombs1970mathematical}.
A higher NMI value indicates a better prediction goodness.

$100$ different topics, spanning over $10$ different Twitter events, were used for experimentation.
The hashtags were chosen for popular events, and each hashtag chosen was unique to a event within the dataset.
For each topic and each event as a whole, significant community overlaps were found between the two graphs, as indicated by NMI values.
Among the unigram, bigram and LDA models, the unigram model was observed to provide the best overlaps.
This is revealing: it suggests significant alikeness of users within social communities, in terms of word (language) usage.
Interestingly, this is in spirit similar to the made by \cite{danescu2013no}, where it is observed that the language usage of social friends evolve and become similar over time (word usage similarity increases).
Further, fine-grained topics found by running LDA on the whole of user content, provided a better approximation (prediction) of the social graph structure, compared to the coarser-grained events represented by hashtags.
The knowledge thus obtained can be used in applications such as social information flow modeling and marketing.

\section{Related Work}
\label{sec:related}

Link prediction on social networks has been an area of long standing research.
\cite{liben2007link} carried out a comprehensive study of different link prediction techniques in a social network setting, including methods such as graph distance, Adamic-Adar method \cite{adamic2003friends}, Jaccard's coefficient \cite{salton1986introduction}, rooted Pagerank \cite{liben2007link}, Katz \cite{katz1953new} and SimRank \cite{jeh2002simrank}.
However, this study, and the subsequent ones in this school of research such as \cite{yin2011structural} and \cite{dong2012link}, focuses on graph structure and properties, and does not consider content semantics.
Subsequently, \cite{puniyani2010social} attempted to study the impact of communication semantics in predicting social links, and used Twitter as their platform for the study.
This study uses LDA \cite{blei2003latent} for predicting pairwise links; however, it neither investigates the relative impact of different language in such prediction, nor does it delve deeper to investigate the social properties of the predicted links, which is essential if one were indeed aiming to predict a {\it social} network.

Researchers have attempted to investigate the flow of information cascades on Twitter, as well as propagation of influence along the underlying social connection graph.
\cite{tsur2012s} predicts the spread of user-generated ideas on Twitter.
\cite{hong2011predicting} proposes a multi-class classification model to identify popular messages on Twitter, by predicting retweet quantities, from TF-IDF (term frequency and inverted document frequency) and LDA, along with social properties of users.
\cite{bakshy2011everyone} models the flow of influence along social connections on Twitter, and makes the surprising observation that in spite of URLs rated interesting and content by influential users spreading more than average, predictions of which particular user or URL will generate large cascades are relatively unreliable.
Other studies, such as \cite{bakshy2012role}, \cite{myers2012information}, \cite{narang2013discovery}, \cite{romero2011differences}, \cite{yang2010predicting} and \cite{yang2010modeling} provide significant insights into flow of information and influence, along social edges, over Twitter user interactions.

In order to find social communities for exploring our problem space, the current work makes use the community finding literature.
The most prominent class of implicit communities formed based upon graph structures is modularity-based communities.
Originally proposed by \cite{newman2006modularity}, a fast approximation algorithm is used, BGLL \cite{blondel2008fast}, to compute max-modularity communities.
In order to derive modularity values, this body of work initially computes the differences of actual and expected (probabilistic) value of a given pair of vertices to have an edge, and subsequently aggregates the above over all possible pairs to maximize the value of modularity.
Cross-entropy \cite{de2005tutorial} is also used, which is based upon Kullback-Leibler (K-L) divergence \cite{kullback1951information}, and normalized mutual information \cite{coombs1970mathematical}, as part of our computation methodology.

Using a combination of social links and user-generated content has been explored in the literature, from different angles.
\cite{ruan2013efficient} attempt to combine the strength of links with similarity of content between each pair of graph vertices (social network users), to augment baseline social link based graphs, and discover communities on such graphs.
\cite{liucommunity} combine the topological structure of a network with the content information present in the network, and thereby model the community structure as a consequence of interaction amongst the participating nodes (social network users).
\cite{yang2013community} also combine the graph node attributes with the graph edge structures for community discovery.
\cite{kuramochi2012community} also consider the overlaps between communities using the concept of the intersection graph, for community discovery.
\cite{natarajan2013community} observe that joint modeling of links and content significantly improves link prediction performance on Twitter subgraphs.

Clearly, there is significant background prior art that exists in related areas.
Some semantic link predictions exist, and some literature also attempt to integrate content with link for community discovery.
However, no direct in-depth investigation exists that attempts to benchmark or compare the goodness of different language models, in predicting or approximating microblog connections while capturing the essential community structures.
Further, no question has ever been raised in the literature on the quality of social structure prediction, or even simple link prediction, with respect to topic identification granularities, with respect to any content or semantic attribute.
We attempt to answer these fundamental questions that have not been investigated in the literature.
This makes our work a first of its kind.

\section{Proposed Approach}
\label{sec:algo}

This section explores the problem settings, and the machinery developed to obtain insights under these settings.

\subsection{Problem Settings}
\label{subsec:probsettings}

The core objective of the current work is to approximate (predict) the microblog connection graph, from user-generated content.
In the process, the following have been proposed.

\begin{itemize}
\item Quantifying the {\it social} goodness of microblog friendship graphs, approximated (predicted) using user-generated content.
\item Investigating the goodness of different language models for the approximation.
\item Studying the impact of granularity of selecting content, to predict the graph structures.
\end{itemize}

\subsection{Solution Framework}
\label{subsec:ourapproach}

\begin{figure*}[thb]
\centering
\includegraphics[scale=0.82]{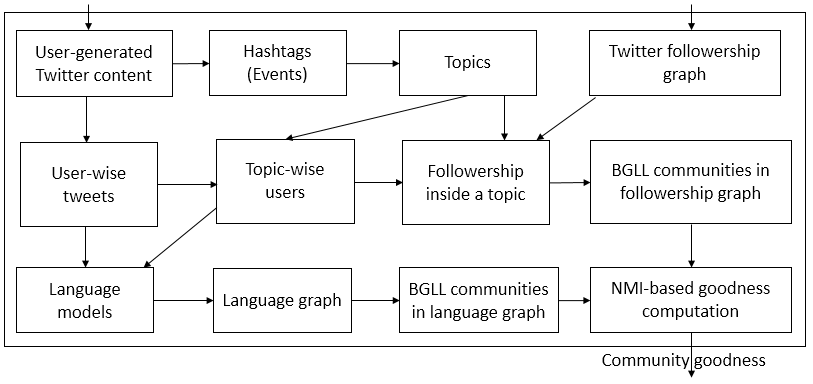}
\caption{The architecture of our framework}
\label{fig:architecture}
\end{figure*}

Our solution framework overlays the structures (communities) formed by two graphs: a graph created by connecting user pairs based upon user-generated content, and a social friendship graph given as ground-truth.
The overlaps of the sets of communities $C_L$ in the content graph and $C_S$ in the social graph are quantified, to obtain insights.
Figure~\ref{fig:architecture} depicts the architecture of our system.

A content graph is first constructed.
The topics present in the document are detected, using LDA.
For each user and each topic, all the tweets that the user makes in her lifetime for the topic are collected.
The set union of these tweets are used to create a document for the user for that topic, as described in Section~\ref{subsec:topicfinding}.

In order to construct semantic edges between user pairs, different language models, such as unigram, bigram and LDA (topic model), are applied.
The techniques to construct these edges (links) are different in case of LDA, from unigram and bigram: the edge construction methodology is explained in Section~\ref{subsec:contentgraph}.
Effectively, these edges are associated with a confidence score, that represents the similarity between user pairs.
The predicted links (constructed edges) are now further processed to extract structural social communities, which is used to evaluate the goodness of the prediction, as detailed in Section~\ref{subsec:structuralgoodness}.
For this, the spectral modularity maximization techniques of Newman \cite{newman2006modularity}, implemented by BGLL \cite{blondel2008fast}, is used.
The ground-truth communities, using known followership (friendship) data present in the social graph, are discovered.
The overlap of these sets of communities are subsequently computed, quantifying using normalized mutual information (NMI).
Higher NMI values indicate better overlap, and hence, better prediction (approximation) quality.
The overview of the approached solution framework shown in Algorithm~\ref{alg:overview}.

\algsetup{indent=2em}
\newcommand{\CAPTION}{\ensuremath{\mbox{\sc Overview of Our Approach}}}
\begin{algorithm*}[htb]
\caption{$\CAPTION$}
\label{alg:overview}

\begin{algorithmic}[1]
\renewcommand{\algorithmicrequire}{\textbf{Input}}
\REQUIRE \textbf{1.} User-generated Twitter content
\REQUIRE \textbf{2.} Twitter followership graph
\newline
\STATE select a set of hashtags (events) from Twitter content
\STATE assign a user $u_i$ to all events (hashtags) that she participates in
\FOR {each event $e$ do}
	\STATE identify topics $c_{e,k}$ within event $e$, using LDA
	\FOR {each user $u_i$ in event $e$ do}
		\STATE find event-level goodness value as the NMI of language and social graph communities
		\STATE construct document $d_i$ as collection of all tweets $t_{e,u_i}$ of $u_i$
		\FOR {each topic $c_{e,k}$}
			\STATE find users $u_{i,e,k}$ participating in topic $c_{e,k}$
			\STATE find language models of these users, using topics and tweets
			\STATE create language graph by drawing edges between user-pairs
			\STATE assign language graph edge weight inversely proportional to user-pair cross entropy
			\STATE identify BGLL communities in the language graph
			\STATE create topic-level social graph of user-pairs, using Twitter followership graph
			\STATE identify BGLL communities in the topic-level social graph
			\STATE find topic-level goodness value by finding NMI of language and social graph communities
		\ENDFOR
	\ENDFOR
\ENDFOR
\newline
\renewcommand{\algorithmicensure}{\textbf{Output:}}
\ENSURE \textbf Set of community goodness scores
\end{algorithmic}
\end{algorithm*}

\section{Implementation}
\label{sec:implementation}
The key steps of the implementation process are described below.

\subsection{Document Creation and Topic Extraction}
\label{subsec:topicfinding}
Documents are generated for each user, as well as the topics present in the content under consideration are extracted.
These topics, and the documents, are subsequently used to create the content graph, by drawing language-based edges between user pairs.

For this, the content generated by each user on the microblog is collected, that contains the hashtag $H_e$.
$H_e$ uniquely identifies event $e$, from a set of events $E$.
Since each tweet is small in length, within 140 characters (typically 14-18 words), the set union of all tweets $t_{eu_i}$ made by user $u_i$, that contain the hashtag $H_e$, is computed.
This forms document $d_{eu_i}$, representing the lifetime participation of user $u_i$ in event $e$, given by Equation~\ref{eqn:peruserdocumentcreate}.
Let $T_e$ denote the complete set of tweets belonging to event $e$.

\begin{equation}
\label{eqn:peruserdocumentcreate}
d_{eu_i} = \mathop{\bigcup}_{(t_{eu_i} \in T_e)} \{t_{eu_i}\}
\end{equation}

Having constructed the individual documents using Equation~\ref{eqn:peruserdocumentcreate}, the topics that are present within the events are extracted.
The overall (broader) document for a given event is extracted, comprising of all tweets having the current event hashtag, as shown in Equation~\ref{eqn:bigdocumentcreate}.

\begin{equation}
\label{eqn:bigdocumentcreate}
D_e = \mathop{\bigcup}_{i} \{d_{ei}\}
\end{equation}

Subsequently, the LDA model is applied, to find the intended topics.
A well-recognized tool, MALLET \cite{mccallum2002mallet}, is used to detect topics $C_e = \{c_{ek}\}$ for a given event $e$, for our experiments.
These topics $c_{ek}$, as well as the per-user documents $d_{ei}$, are subsequently used for creating the content graph, and thereby for modeling the weighted semantic links between pairs of users.

\subsection{Content Graph Creation}
\label{subsec:contentgraph}
Creating the content graph involves identifying content creation similarity among user pairs, measured by a chosen language model, and thereby identifying links between the pair.
Two classes of language models are selected: (a) the unigram and bigram models from the $n$-gram language model class, and (b) the LDA model from the probabilistic semantic model class.\\

\noindent \textbf{Edge creation in n-gram models}\\
The n-gram language model $l_i$ is computed for each user $u_i$ that have participated in any of the events, and any topic of the event.
It should be noted that the language model is independent of any event or topic that the user participates in.
For each given event $e$, edges between each pair of user participating in that event are created.
Cross-entropy \cite{de2005tutorial} being inherently asymmetric, to create an edge between a given pair of users $u_{e,i_1}$ and $u_{e,i_2}$, we take the average of cross-entropy of the language model $l_1$ of user $u_{e,u_1}$ with respect to the other user's event-level document $d_{e,u_2}$, and that of $l_2$ w.r.t. $d_{e,u_1}$.
Weights are assigned to the edges as the difference of the maximum value of the cross-entropy and the value of the cross-entropy of the current edge (since high cross-entropy denotes low similarity).
Thresholds are subsequently applied, to finally retain or discard the edge created hereby, by selecting the top $k\%$ edges for experimentation.

Apart from constructing the edges for the event-level graphs, constructing edges for the topic-level graphs is also necessary, for the topics that were found by the LDA model within each event.
For for each document $d_{e,u_i}$, we find the probability $p_{e,d_{e,u_i},c_{ek}}$ of the document to belong to a topic $c_{e,k}$.
If there are $K$ topics for the event $e$, then is it assumed that a user is associated with the topic as long as $p_{e,d_{e,u_i},c_{ek}} > \frac{1}{K}$.
The intuition behind this is simple: it retains all the topics in which the user's participation was higher than the expected random participation, denoting user interest in the topic.
This will create a subset of users $u_i$, that participate in event $e$.
The cross-entropy based edge-creation process is repeated at the topic level also.
For each topic, weight inversely proportional to the cross-entropy values is assigned.
This creates one graph per topic, within a given event.

Threshold-based retention of the event-level or topic-level edges, generated by the above, is subsequently applied.
For experimentation, instead of considering any absolute threshold (it is scientifically difficult to establish a {\it good} threshold), the goodness of our system is tested by retaining a certain percentage of the best edges (high values of weights), and discarding the rest.
This is useful because it provides an intuition into optimally selecting (predicting) edges that can be easily identified, rather than all edges, for our current purposes.

It should be noted that, if $H(P)$ denotes the marginal entropy of a probability distribution $P$, the cross-entropy of a probability distribution $Q$ w.r.t. $P$ is given as:

\begin{equation}
\label{eqn:crossentropy}
H(P,Q) = H(P) + D_{KL}(P||Q)
\end{equation}

Here, $D_{KL}(P||Q)$ is the K-L divergence \cite{kullback1951information} of $Q$, a probability distribution, w.r.t. $P$, another probability distribution.
Further, for a discrete distribution such as ours, K-L divergence is given by

\begin{equation}
\label{eqn:kldivergence}
D_{KL}(P||Q) = \sum\limits_{i}{} P(i) ln\frac{P(i)}{Q(i)}
\end{equation}

The n-gram model based content edge creation process is applied for $n=1$ (unigrams) and $n=2$ (bigrams).
The n-gram probability distribution $Q$, and the document as $P$, are used to compute cross-entropy. \\

\noindent \textbf{Edge creation in the LDA model}\\
In order to create the edges in the LDA model, the probability $q_{e,d_{e,u_i},c_{ek}}$ is computed for each user $u_i$, for participating in each topic, by assigning the user's document $d_{e,u_i}$ a probability to belong to each topic $c_{e,k}$.
If there are $K$ topics for the event $e$, then it is assumed that a user is associated with the topic as long as $q_{e,d_{e,u_i},c_{ek}} > \frac{1}{K}$.
This creates a subset of users of the event, to participate in the topic.
An event-level edge between a given user pair, say $u_1$ and $u_2$, with weight computed as a sum-of-product of the probabilities (Equation~\ref{eqn:eventedges}), is created.

\begin{equation}
\label{eqn:eventedges}
w^{LDA}_{u_1,u_2} = \sum\limits_{k}{} (q_{e,d_{e,u_1},c_{ek}} * q_{e,d_{e,u_2},c_{ek}})
\end{equation}

In case of a topic model, a similar process is applied on the subset of users participating in the topic.
Subsequently, a threshold-based edge retention process is carried out, in a manner similar to what was done in the n-gram model.
This completes the process of constructing the content graph (effectively, prediction of the links).
The three sets of graphs derived from the three language/topic models, namely unigram, bigram and LDA, and each constructed at two granularities, namely event and topic levels, are now assessed for structural goodness.

\subsection{Measuring Structural Goodness}
\label{subsec:structuralgoodness}
Most of the related literature, including the LDA-based edge prediction by \cite{puniyani2010social}, has measured the goodness of their work, using prediction accuracies and error rates.
On the contrary, the current work aims to measure the {\it social} goodness of the predicted links.
Therefore, the overlap of the communities (social structures) formed by the predicted links (semantic graph), with the ground truth (social graph), is quantified.
Thus, the prediction is effectively evaluated at a structure level, which has much deeper social semantics compared to individual links.

This is done in two stages.
First, the modularity-based structural communities in the content graph, as well as the social friendship graph, are independently detected.
Subsequently, the overlap two community sets is measured, by computing the NMI within each of these two sets.
By definition \cite{coombs1970mathematical}, NMI values range between 0 and 1, and a higher NMI value indicates a higher overlap of the two sets of communities.\\

\noindent \textbf{Finding modularity communities}\\
The concept of modularity \cite{newman2006modularity} is used to discover implicit structural communities.
Modularity technique aims to partition a given graph into non-overlapping components, maximizing the proportion of connections between pairs of vertices belonging to the same component, to that of pairs of vertices belonging to two different components.
To compute modularity, two quantities are considered for a given pair of components.
(a) The ground truth, which is deterministic about whether a given pair of vertices are connected by an edge, or not.
(b) The probabilistic expectation that a given pair of vertices is connected, given the total degree and total number of vertices of the entire graph.
The partitioning is carried out by computing the differences from the former with the latter, and aggregating over the arrangements.

Newman's spectral method for computing modularity is formulated as:
\begin{equation}
\label{eqn:modularity}
Q = \frac{1}{4m} \sum\limits_{i,j}(A_{ij} - \frac{k_ik_j}{2m})s_is_j
\end{equation}

Here $Q$ denotes the modularity, $A_{ij}$ are the adjacency matrix elements (edge) between vertices $i$ and $j$, $k_ik_j/2m$ is the expected number of edges between vertices $i$ and $j$ when placed at random, $1/4m$ is a conventional factor and $s_i$ and $s_j$ are components (communities) that vertices $i$ and $j$ belong to.
Newman's method is computationally expensive, taking $O((m+n)n)$ time, where $n$ is the number of vertices and $m$ is the number of edges in the graph.
Instead, the BGLL \cite{blondel2008fast} algorithm is used, since it provides a fast implementation of modularity computation.\\

\noindent \textbf{Computing NMI}\\
Mutual information is computed as
$$I(X,Y)=\sum_{y \in Y}\sum_{x \in X}p(x,y)log(\frac{p(x,y)}{p(x)p(y)})$$
where $I(X,Y)$ is the mutual information of $X$ and $Y$.
The normalized mutual information (NMI) \cite{coombs1970mathematical} across these two sets of communities is computed as
$$C_{XY}=\frac{I(X,Y)}{H(Y)}$$
and
$$C_{YX}=\frac{I(X,Y)}{H(X)}$$
respectively, where H(X) and H(Y) denote the marginal entropies of X and Y respectively.

\section{Experiments}
\label{sec:expt}
Experiments were carried out on data from $10$ Twitter events, having unique hashtags.
Further, $10$ LDA-based topics were identified within each event using MALLET \cite{mccallum2002mallet}.
We thus validate our hypotheses experimentally, over $10$ events and $10 * 10 = 100$ topics.
The tools used were: MALLET, to find topics and find the LDA-based content model of users, and statistical language modeling toolkit by \cite{rosenfeld1997statistical} for unigram and bigram models, and cross-entropy.

\subsection{Data Description}
\label{subsec:datadescription}
Twitter data was collected over $10$ different events, where each event was identified by a unique hashtag.
The following facets of the data were collected:
(a) For each hashtag, all data (content) that was generated.
(b) All tweets (content) that each user, who ever posted with a given hashtag, in their lifetime on Twitter (limited by Twitter max of $3,200$).
(c) Followership graph of each of these users.
Implicit reciprocity in the followership graph is assumed. 
For the sake of brevity, we present the experimental results for $60$ topics, spanning over $6$ randomly chosen events from the $10$.
Table~\ref{tab:basicstats} presents basic statistics of these events.

\begin{table}[htb]
\caption{Basic statistics of event datasets. Edges are based on followership, ignoring directions.}
\begin{center}
\begin{tabular}{|r|r|r|r|}
\hline
\textbf{Event} & \textbf{Num} & \textbf{Num} & \textbf{Num} \\
\textbf{Hashtag} & \textbf{Nodes} & \textbf{Edges} & \textbf{Tweets} \\
\hline
Billboards & $327$ & $778$ & 7,579 \\
Cesar & $1,005$ & $6,013$ & 5,273 \\
Coachella & $453$ & $604$ & 8,876 \\
Elections & $473$ & $1,575$ & 8,815 \\
Junos & $624$ & $3,366$  & 4,334 \\
Ted & $631$ & $1,243$ & 7,184 \\
\hline
\end{tabular}
\end{center}
\label{tab:basicstats}
\end{table}

\subsection{Evaluating Content Models}
\label{subsec:contentmodeleval}
\begin{figure}[tbh]
\centering
	\subfigure[Events: Billboards, Cesar]{
		\includegraphics[width=0.4\textwidth]{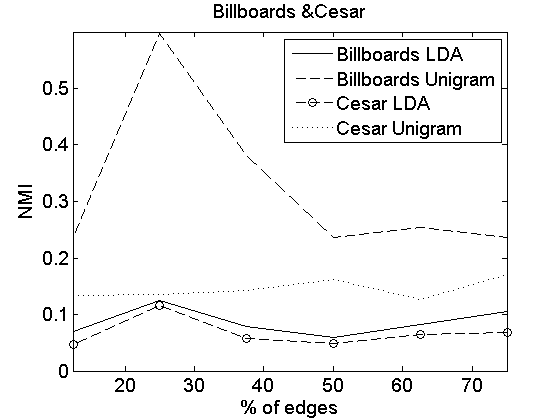}
		\label{fig:lang_billboards_cesar}
	}
	\subfigure[Events: Coachella, Elections]{
		\includegraphics[width=0.4\textwidth]{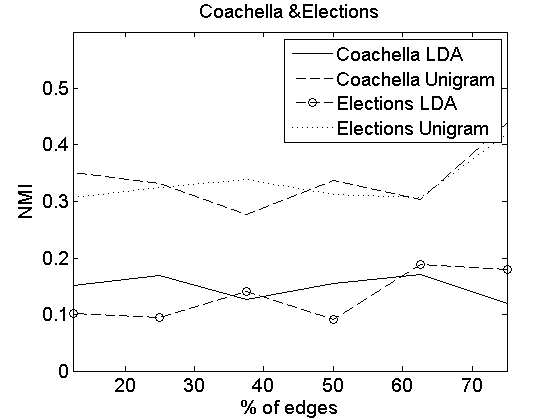}
		\label{fig:lang_coachella_elections}
	}
	\subfigure[Events: JUNOS, TED]{
		\includegraphics[width=0.4\textwidth]{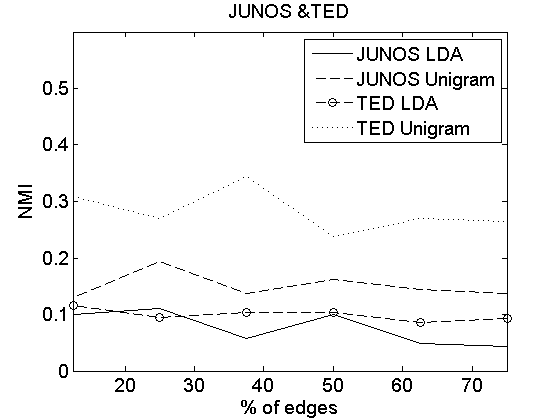}
		\label{fig:lang_junos_ted}
	}
\caption{Effectiveness of content models as approximators of social structures (communities). $2$ hashtags covered per plot.} 
\label{fig:contentmodels}
\end{figure}

Content models, namely unigram and bigram language models and LDA topic model, were evaluated at this stage.
Figure~\ref{fig:contentmodels} shows the variation of NMI for each content model, at different threshold levels, across multiple events.
The results obtained with bigrams were below par, hence the experiments presented were conducted with unigram and LDA.
The threshold levels have been chosen to retain the top $K\%$ of the edges (ranked by weights).

\begin{figure*}[thb]
\centering
	\subfigure[Event: Elections]{
		\includegraphics[width=0.4\textwidth]{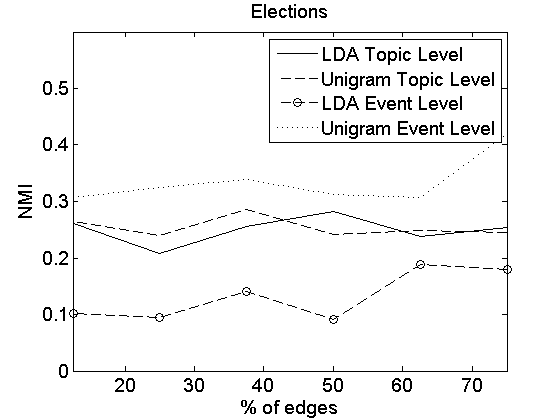}
		\label{fig:gran_elections}
	}
	\subfigure[Event: JUNOS]{
		\includegraphics[width=0.4\textwidth]{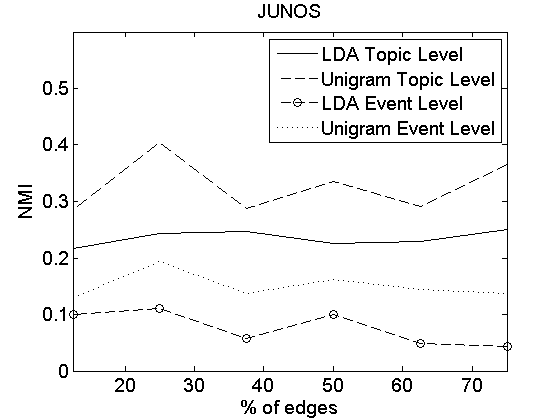}
		\label{fig:gran_junos}
	}
	\subfigure[Event: TED]{
		\includegraphics[width=0.4\textwidth]{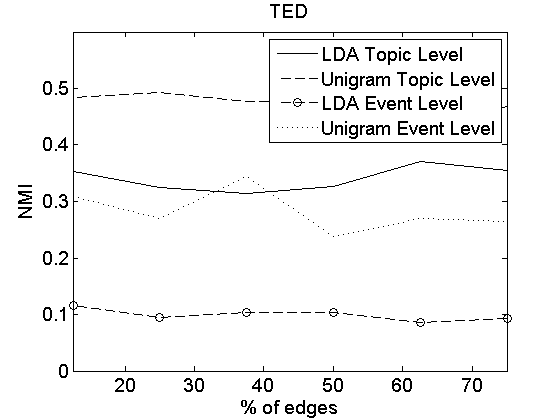}
		\label{fig:gran_ted}
	}
	\subfigure[Event: Coachella]{
		\includegraphics[width=0.4\textwidth]{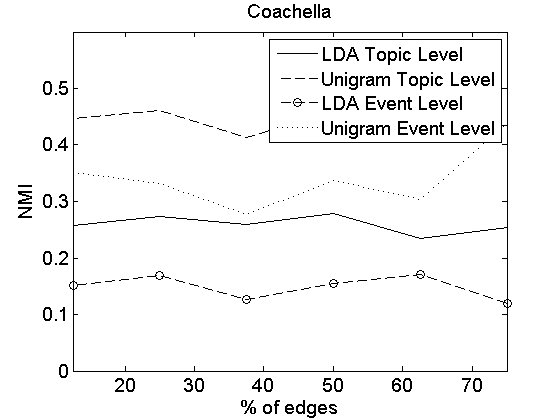}
		\label{fig:gran_coachella}
	}
	\subfigure[Event: Cesar]{
		\includegraphics[width=0.4\textwidth]{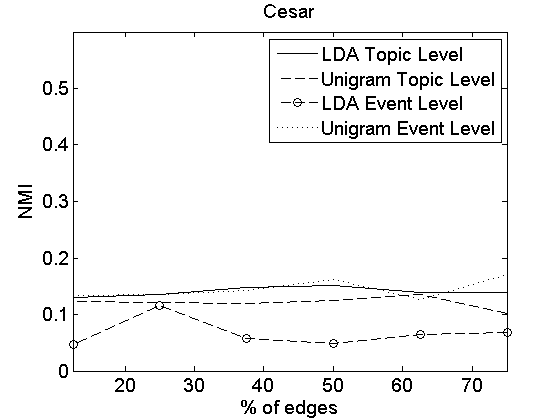}
		\label{fig:gran_cesar}
	}
	\subfigure[Event: BillBoards]{
		\includegraphics[width=0.4\textwidth]{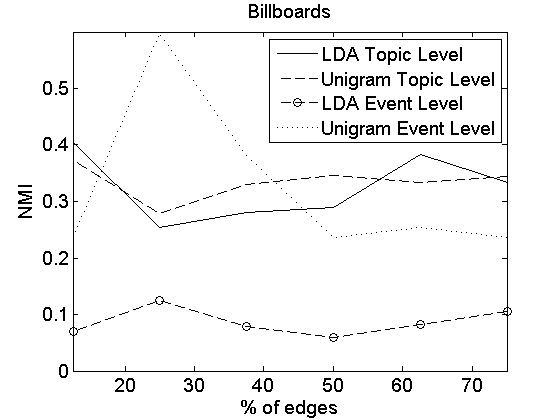}
		\label{fig:gran_billboards}
	}
\caption{Impact of corpus granularity selection on NMI, at topic (fine) vs. event (coarse) levels.}
\label{fig:granularity}
\end{figure*}

In our experiments, the unigram models performed better than LDA consistently across all the datasets.
This is clear from the NMI values, which are higher for unigram, compared to LDA.
Plots for $6$ events are shown in Figure~\ref{fig:granularity}; however, these trends were consistently observed for all the events and topics explored.
This indicates significant similarity of word usage within social communities, which is well-captured by unigram, but given theme (topic) similarity would create confusions in case of LDA.
Manual inspection of ground-truth friendship edges, specifically predicted by unigram but not by LDA, confirm these trends.

\subsection{Evaluating Granularities}
\label{subsec:granularityeval}
The impact of granularity of the chosen corpus, namely the coarser-grained event and finer-grained topic level granularity, on NMI values, was evaluated over different thresholds.
Figure~\ref{fig:granularity} demonstrates the trends.
Finer-grained topics consistently yielded higher NMI values, compared to coarser-level events.
This shows the effectiveness of selecting finer-grained topic-level tweets as corpus to predict social structures, over coarser-grained event-level tweets.

\subsection{Observing Community Structures}
\label{subsec:commstruct}
Our experiments revealed significant NMI values on the content based graph communities, with respect to the explicit friendship based social communities.
Since friendship connections are ground truth, the NMI values indicate that the content models have been able to retain much of the communities, a core social property, in the predicted links.
This demonstrates the effectiveness of content models, in predicting ``good'' social links from user-generated text in Twitter, with the unigram model being the most effective (validated by retaining the community structures better than the other models, for all the datasets).
Finally, finer-grained topics were observed to be better approximators of social communities, compared to coarser-grained events.

\section{Discussion}
\label{subsec:discussions}
The current study elicits a surprising insight: in spite of being inherently a much simpler model compared to LDA, the unigram model was seen to be more effective in capturing the links that are better in attaining a deep and complex social property - social communities.
This can be attributed to word usage behavior similarity within social communities (exact same words getting used within communities), favoring the unigram model, leading to spurious edges in LDA.

The authors have noted that in recent literature, the language usage behavior of individuals has been shown to evolve and become similar to other individuals belonging to the same communities, over time \cite{danescu2013no}.
\cite{danescu2013no} exemplify with {\it aroma} and {\it smell}: they observe that communities built upon the keyword {\it beer}, at one point of time, tended to use the word {\it aroma} together, which over time evolved into {\it smell} (or, S in short).
However, the phenomenon of using the same word (effectively, {\it unigram similarity}) was observed among the wide span of connected individuals belonging to the communities, not just at the level of individuals.
Note that, this observation is strongly in alignment with our observations of unigram outperforming LDA: LDA would capture {\it smell} and {\it aroma} to be similar, and use that as a predicting feature of a social connection, while unigram will not.
The unigram match outperforming other content models is likely to be a reflection of this very phenomenon.
In other words, while LDA gives a probabilistic distribution of language usage for individuals, unigrams, the specific words used by individuals, match with each other as individuals belonging to the same community pick up word usage behavior of others and tend to start using similar words.
This is an interesting observation about the deeper language usage behavior of individuals belonging to similar communities.

Thus, while much of the current literature adapt LDA for predicting content-based social links, we are the first ones in the literature to raise a question regarding whether, in spite of its richness, LDA is at all the most effective content model, and surprisingly observe on the contrary.
Further, since our work captures the tweets made by each user under consideration for their entire lifetime on Twitter that is made publicly available, our LDA model is trained on as much Twitter data that one can access, and what every other content based link construction is made from in the literature, that uses the public Twitter APIs.

\section{Conclusions}
\label{sec:concl}
In the current work, a framework was created to study the effectiveness of language models in approximating or predicting microblog connection structures.
Hashtags were used to identify coarse-grained events.
LDA-based fine-grained topics were found within each event.
The participation of users were found, in each topic within each event, from a given set of events, based upon user generated content.
Using the language usage similarities and topic similarities of all user pairs, a content-based user graph was created, spanning participants of the event-related discussions.
Unigram, bigram and LDA were used as the underlying models.
For each event, the overlap of the language graph structure, with the ground-truth social graph structure, was quantified at different content granularities such as event-level and topic-level, using NMI.
Experiments were conducted with $100$ topics, spanning over $10$ Twitter events, for empirically proving our proposition.
The results consistently demonstrate the goodness of the approximation, at different granularities, with higher NMI values emanating from more fine-grained topics.
The unigram model was consistently found to be most effective, in all the cases.
This indicates a strong similarity of word usage behavior of users within deeply connected social communities.
Some applications of the current work would be in the academic area of information flow modeling, as well as practical field of social marketing based applications.

\bibliographystyle{IEEEtran}
\bibliography{IEEEabrv,arxiv}
\end{document}